\documentclass[
  aps,
  prd,
  reprint,
  footinbib,
  superscriptaddress,
  floatfix
]{revtex4-2}

\usepackage{amsmath,amssymb,amsthm,mathtools,bm}
\usepackage{hyperref}
\usepackage{microtype}

\begin{document}

\title{Static axisymmetric Einstein spaces with a cosmological constant and the limitation of canonical Weyl coordinates}

\author{Sheref Nasereldin}
\affiliation{Centre for Theoretical Physics, The British University in Egypt, P.O. Box 43, El Sherouk City, Cairo 11837, Egypt}
\affiliation{Universities of Canada in Egypt, New Administrative Capital City, Plot No. (X1-05), Cairo, Egypt}

\date{\today}

\begin{abstract}
The canonical Weyl form for four-dimensional static axisymmetric vacuum metrics is obtained by identifying the area function of the two Killing orbits with a harmonic coordinate on the two-dimensional orbit space. This construction is valid in Ricci-flat vacuum, but it is no longer available in Einstein spaces with nonzero cosmological constant. In this paper, we consider the generalized orthogonally transitive static axisymmetric line element and derive the reduced Einstein--$\Lambda$ field equations. We show that the canonical Weyl choice $W=\rho$ is locally admissible if and only if $\Lambda=0$. The Kottler metric gives the simplest explicit example of the resulting equation for the area function. Thus, the statement that ``Weyl metrics do not allow $\Lambda\neq0$'' is precise only when the metric is assumed to be in canonical Weyl coordinates. The issue is not staticity or axisymmetry, but rather the fact that the area function is no longer harmonic.
\end{abstract}

\maketitle

\section{Introduction}

The classical Weyl metric gives the general local form of the exterior vacuum field of a static axially symmetric gravitating source in four dimensions \cite{Weyl1919}. It is the static specialization of the Weyl--Lewis--Papapetrou construction for spacetimes with two commuting Killing fields \cite{Lewis1932,Papapetrou1966,StephaniBook}. In canonical coordinates its line element reads
\begin{equation}\label{ansatz}
  ds^2 = - e^{2U} dt^2 + e^{-2U}\Bigl[e^{2\gamma}(d\rho^2 + dz^2) + \rho^2 d\phi^2\Bigr],
\end{equation}
where $U = U(\rho,z)$ and $\gamma = \gamma(\rho,z)$. In vacuum, $U$ satisfies the flat-space axisymmetric Laplace equation and $\gamma$ is obtained by quadratures; for example, see \cite{GriffithsPodolsky2009}.

Static spacetimes with nonzero cosmological constant also exist, with the Kottler family \cite{Kottler1918} being the simplest example. The question now arises as to whether these Einstein spaces can be put in the same canonical Weyl form.

It is important to note that the Weyl line element is not merely a statement about staticity and axial symmetry. It also imposes the canonical Weyl coordinate choice
\begin{equation}
  W \equiv \sqrt{-\det(g_{AB})} = \rho, \qquad A,B \in \{t,\phi\}.
\end{equation}
For Ricci-flat vacuum, $W$ is harmonic on the orbit space, and this coordinate choice is admissible. For Einstein spaces with
\begin{equation}
  R_{ab} = \Lambda g_{ab},
\end{equation}
the same function satisfies an equation with a nonzero source term. Therefore, the limitation is not staticity itself, but the use of canonical Weyl coordinates.

The purpose of this paper is to make this statement precise in the purely static axisymmetric setting. We work locally in the orthogonally transitive sector, derive the reduced Einstein--$\Lambda$ field equations for a generalized line element with area function $W(\rho,z)$, and show that the canonical choice $W=\rho$ is locally admissible in a region away from the symmetry axis, $\rho>0$, if and only if $\Lambda=0$. The Kottler metric then provides the simplest explicit illustration of the resulting equation for $W$.

This question is also connected to earlier work on reductions of Einstein spaces with a cosmological constant and on $\Lambda$-deformed integrability \cite{LeighEtAl2014,KlemmNozawaRabbiosi2015,CardosoMayorgaPenaNampuri2024,CardosoMahapatraNagy2024}. The point emphasized here is more specific: in the static axisymmetric sector, the failure of the canonical Weyl form is controlled directly by the differential equation satisfied by the area function. Thus, the result should be viewed as a local statement about canonical Weyl coordinates, not as a global obstruction to static axisymmetric Einstein spaces.

\section{Static axisymmetry beyond canonical Weyl coordinates}

To keep the canonical Weyl condition out of the ansatz (\ref{ansatz}), we work with the more general orthogonally transitive static axisymmetric line element
\begin{equation}
  ds^2
  =
  - e^{2U} dt^2
  + e^{-2U}\left[e^{2\nu}(d\rho^2 + dz^2) + W(\rho,z)^2 d\phi^2\right].
  \label{eq:general-static-axisym}
\end{equation}
Here $U=U(\rho,z)$, $\nu=\nu(\rho,z)$, and $W=W(\rho,z)$. This is the usual local setting in which the Weyl--Lewis--Papapetrou form is introduced \cite{Papapetrou1966,StephaniBook}. The canonical Weyl metric is recovered by the additional specialization
\begin{equation}
  W(\rho,z) = \rho.
\end{equation}
This separates the coordinate specialization from the broader existence of static axisymmetric Einstein spaces.

\section{Reduced Einstein--\texorpdfstring{$\Lambda$}{Lambda} equations}

The line element \eqref{eq:general-static-axisym} leaves the area function $W(\rho,z)$ arbitrary. We now use the Einstein--$\Lambda$ field equations to determine whether the canonical Weyl choice $W=\rho$ is compatible with the field equations, or whether it is a special feature of the Ricci-flat case.

We work under the assumption of staticity. Namely, there exists a timelike Killing vector $\xi^\alpha=\delta^\alpha_t$ satisfying
\begin{equation}
  \xi_{[\alpha}\nabla_\beta\xi_{\delta]}=0,
\end{equation}
and all metric functions are independent of $t$. In this case, the line element can be written in the standard static form \cite{Wald1984,StephaniBook}
\begin{equation}
  ds^2 = - e^{2U(\rho,z)} dt^2 + h_{ij}(x)\, dx^i dx^j,
\end{equation}
where $x^i=(\rho,z,\phi)$, and $h_{ij}$ is the metric induced on the hypersurfaces orthogonal to $\xi^\alpha$.

If $R_{\mu\nu} = \Lambda g_{\mu\nu}$, then the lapse potential $U$ and the spatial metric $h$ satisfy \cite{Note1}
\begin{align}
  \Delta_h U + |DU|_h^2 &= - \Lambda,
  \label{eq:static-einstein-1} \\
  \mathrm{Ric}(h)_{ij}
    &= D_i D_j U + D_i U D_j U + \Lambda h_{ij}.
  \label{eq:static-einstein-2}
\end{align}
where $D$ and $\Delta_h$ denote the Levi-Civita connection and Laplacian associated with the spatial metric $h$, respectively.

We now make the conformal rescaling
\begin{equation}
  h_{ij} = e^{-2U}\,\bar h_{ij}.
\end{equation}
With this choice, the four-metric takes the Weyl-like form
\[
  ds^2 = -e^{2U}dt^2 + e^{-2U}\bar h_{ij}dx^i dx^j .
\]
In the orthogonally transitive axisymmetric setting, the rescaled three-metric can then be written as
\begin{equation}
  \bar h = q_{AB}(x)\, dx^A dx^B + W^2(x)\, d\phi^2,
  \qquad A,B \in \{\rho,z\},
\end{equation}
where $q$ is the metric on the two-dimensional orbit space and $W$ measures the size of the axial circles. Finally, choosing local conformal coordinates $(\rho,z)$ on the orbit space gives
\begin{equation}
  q_{AB}dx^A dx^B = e^{2\nu(\rho,z)}(d\rho^2 + dz^2).
\end{equation}

By virtue of the conformal transformation formulas in three dimensions, \eqref{eq:static-einstein-1} and \eqref{eq:static-einstein-2} become
\begin{align}
  \Delta_{\bar h} U &= - \Lambda e^{-2U},
  \label{eq:barh-laplacian} \\
  \mathrm{Ric}(\bar h)_{ij}
    &= 2 \, \partial_i U \, \partial_j U + 2 \Lambda e^{-2U} \bar h_{ij}.
  \label{eq:barh-ricci}
\end{align}

For the three-metric
\begin{equation}
  \bar h = q_{AB} dx^A dx^B + W^2 d\phi^2,
\end{equation}
the nonzero Ricci components are
\begin{align}
  \mathrm{Ric}(\bar h)_{AB} &= \mathrm{Ric}(q)_{AB} - W^{-1}\nabla_A \nabla_B W, \\
  \mathrm{Ric}(\bar h)_{\phi\phi} &= - W \Delta_q W,
\end{align}
where all derivatives are taken with respect to $q$.

Substituting these expressions into the conformally rescaled static equations gives the reduced Einstein--$\Lambda$ system on the two-dimensional orbit space:
\begin{align}
  \Delta_q U + W^{-1}\nabla W \cdot \nabla U &= - \Lambda e^{-2U},
  \label{eq:reduced-U-invariant} \\
  \Delta_q W &= - 2 \Lambda e^{-2U} W,
  \label{eq:reduced-W-invariant} \\
  \mathrm{Ric}(q)_{AB}
    &= W^{-1}\nabla_A \nabla_B W + 2 \nabla_A U \nabla_B U \nonumber \\
    &\quad + 2 \Lambda e^{-2U} q_{AB}.
  \label{eq:reduced-q-invariant}
\end{align}

The equation for $W$ comes from the $\phi\phi$ component of \eqref{eq:barh-ricci}. Since $U$ is independent of $\phi$, \eqref{eq:barh-laplacian} gives
\begin{equation}
  \Delta_{\bar h} U = \Delta_q U + W^{-1}\nabla W \cdot \nabla U,
\end{equation}
which gives the equation for $U$. The remaining $AB$ components of \eqref{eq:barh-ricci} give the tensor equation for $q$.

A compact derivation of \eqref{eq:static-einstein-1}--\eqref{eq:reduced-q-invariant} is given in the Appendix.

In the conformal coordinates $q_{AB} = e^{2\nu}\delta_{AB}$, the reduced system becomes
\begin{widetext}
\begin{align}
  U_{,\rho\rho} + U_{,zz}
    + \frac{W_{,\rho}U_{,\rho} + W_{,z}U_{,z}}{W}
    &= - \Lambda e^{2\nu-2U},
  \label{eq:U-coordinate} \\
  W_{,\rho\rho} + W_{,zz}
    &= - 2 \Lambda e^{2\nu-2U} W,
  \label{eq:W-coordinate} \\
  \nu_{,\rho\rho} + \nu_{,zz}
    &= - \bigl(U_{,\rho}^2 + U_{,z}^2\bigr) - \Lambda e^{2\nu-2U}.
  \label{eq:nu-trace}
\end{align}
The off-diagonal and anisotropic parts of \eqref{eq:reduced-q-invariant}
give the following first-order relations:
\begin{align}
  W_{,\rho}\nu_{,\rho} - W_{,z}\nu_{,z}
    &= W\bigl(U_{,\rho}^2 - U_{,z}^2\bigr)
    + \frac{1}{2}\bigl(W_{,\rho\rho} - W_{,zz}\bigr),
  \label{eq:nu-first-rho} \\
  W_{,z}\nu_{,\rho} + W_{,\rho}\nu_{,z}
    &= 2W U_{,\rho}U_{,z} + W_{,\rho z}.
  \label{eq:nu-first-z}
\end{align}
\end{widetext}

\section{Relation with the canonical Weyl identity}

For the canonical Weyl metric one has
\begin{equation}
  g_{\rho\rho} = g_{zz}.
\end{equation}
In addition, the Einstein tensor obeys the identity
\begin{equation}
  G_{\rho\rho} = - G_{zz},
  \label{eq:GrrminusGzz}
\end{equation}
which is the content behind the footnote in \cite{NasereldinLake2019}. This provides a useful check on the more general reduction derived above.

If one now imposes the Einstein--$\Lambda$ equations,
\begin{equation}
  G_{ab} + \Lambda g_{ab} = 0,
\end{equation}
then, by virtue of $g_{\rho\rho} = g_{zz}$, we obtain
\begin{equation}
  G_{\rho\rho} = G_{zz}.
\end{equation}
Combining this with \eqref{eq:GrrminusGzz} yields
\begin{equation}
  G_{\rho\rho} = G_{zz} = 0,
\end{equation}
and hence necessarily $\Lambda = 0$.

The same conclusion follows directly from the reduced equation for $W$. If one imposes the canonical Weyl coordinate choice $W=\rho$, \eqref{eq:W-coordinate} reduces to
\begin{equation}
  0 = - 2 \Lambda e^{2\nu-2U}\rho.
\end{equation}
Away from the symmetry axis, $\rho>0$, the exponential factor is nonzero, and therefore $\Lambda=0$. Conversely, when $\Lambda=0$, \eqref{eq:W-coordinate} becomes the flat two-dimensional Laplace equation for $W$, so the canonical Weyl coordinates are locally admissible wherever $\nabla W \neq 0$. Thus, within the orthogonally transitive static axisymmetric line element \eqref{eq:general-static-axisym}, the canonical Weyl choice $W=\rho$ is compatible with the field equations in a local region away from the axis if and only if $\Lambda=0$.

\section{The area function and the correct replacement}

Define the area function of the Killing orbits by
\begin{equation}
  W = \sqrt{-\det(g_{AB})}, \qquad A,B \in \{t,\phi\}.
\end{equation}
For the general line element \eqref{eq:general-static-axisym}, this is simply the metric function $W(\rho,z)$.
This definition is useful because it is tied to the two Killing directions rather than to a particular choice of the orbit-space coordinates. Under a constant nonsingular linear redefinition of the Killing coordinates, $W$ changes only by the absolute value of the determinant of that linear transformation. Therefore, the question of whether $W$ is harmonic, and hence whether it can be used as a canonical Weyl coordinate up to a constant normalization, is not an artifact of the coordinates $(\rho,z)$.

In Ricci-flat vacuum, $W$ is harmonic on the two-dimensional orbit space. This enables one to choose $W$ itself as a coordinate and obtain the canonical Weyl form. For $\Lambda \neq 0$, $W$ satisfies an equation with a nonzero source term, so there is no reason to expect the coordinate choice $W=\rho$ to remain admissible.

\section{Kottler as the simplest example}

Consider the Kottler metric \cite{Kottler1918}
\begin{equation}
  ds^2 = - f(r) dt^2 + f(r)^{-1} dr^2 + r^2(d\theta^2 + \sin^2\theta \, d\phi^2),
\end{equation}
with
\begin{equation}
  f(r) = 1 - \frac{2M}{r} - \frac{\Lambda r^2}{3}.
\end{equation}
This metric is static and axisymmetric, so it must be contained in \eqref{eq:general-static-axisym}. One reads off
\begin{equation}
  e^{2U} = f(r),
  \qquad
  W = r \sin\theta \sqrt{f(r)}.
\end{equation}
The two-metric $q$ appearing in \eqref{eq:general-static-axisym} is
\begin{equation}
  q = dr^2 + fr^2 d\theta^2.
\end{equation}

A direct computation gives
\begin{equation}
  \Delta_q W
  =
  \frac{\sqrt{f}\sin\theta}{r}
  \left(
    f + 2 r f' + \frac{r^2}{2} f'' - 1
  \right),
\end{equation}
and substituting the Kottler expression for $f$ yields
\begin{equation}
  \Delta_q W = - 2 \Lambda e^{-2U} W,
\end{equation}
which is in exact agreement with \eqref{eq:reduced-W-invariant}.

Thus, when $\Lambda = 0$, $W$ is harmonic and canonical Weyl coordinates are admissible. In contrast, when $\Lambda \neq 0$, $W$ is not harmonic and the canonical Weyl coordinate choice fails.
In particular, the $\Lambda=0$ member of the Kottler family is the Schwarzschild solution. In that case, the same expression for $W$ satisfies $\Delta_q W=0$, which is the familiar condition that permits the passage to canonical Weyl coordinates.

\section{Conclusion}

The canonical Weyl form is a special feature of the Ricci-flat static axisymmetric sector. For Einstein spaces with nonzero cosmological constant, the natural object to track is the area function $W$, and the reduced field equations show that $W$ satisfies an equation with a source term rather than a harmonic equation.

The reduced system \eqref{eq:reduced-U-invariant}--\eqref{eq:reduced-q-invariant} makes this statement explicit. In particular, \eqref{eq:reduced-W-invariant} implies that the canonical choice $W=\rho$ is locally compatible with the field equations if and only if $\Lambda=0$. The Kottler metric provides the simplest concrete realization of this result.

Accordingly, the statement that Weyl metrics do not allow a nonzero cosmological constant is correct only when ``Weyl metric'' is understood in the canonical sense, namely, with the area function itself identified with the coordinate $\rho$. Static axisymmetric Einstein spaces with $\Lambda \neq 0$ do exist, but they lie outside that coordinate choice and must be described locally by the more general line element \eqref{eq:general-static-axisym}.

\section*{Data Availability Statement}

No new data were created or analyzed in this study.

\appendix

\section{Reduction formulas}

This appendix records the identities used in Sec.~III.

\subsection{Static Einstein equations}

Let
\begin{equation}
  ds^2 = -N^2 dt^2 + h_{ij} dx^i dx^j,
  \qquad N = e^U,
\end{equation}
with all metric functions independent of $t$. For an Einstein space,
\begin{equation}
  R_{\mu\nu} = \Lambda g_{\mu\nu},
\end{equation}
the standard static decomposition gives
\begin{align}
  R_{tt} &= N D^i D_i N = - \Lambda N^2, \\
  R_{ij}(g) &= \mathrm{Ric}(h)_{ij} - N^{-1} D_i D_j N = \Lambda h_{ij}.
\end{align}
Since $N=e^U$,
\begin{align}
  N^{-1}D^iD_iN &= \Delta_h U + |DU|_h^2, \\
  N^{-1}D_iD_jN &= D_iD_jU + D_iU D_jU,
\end{align}
and hence
\begin{align}
  \Delta_h U + |DU|_h^2 &= - \Lambda, \\
  \mathrm{Ric}(h)_{ij} &= D_iD_jU + D_iU D_jU + \Lambda h_{ij}.
\end{align}

\subsection{Conformal rescaling}

Now write
\begin{equation}
  h_{ij} = e^{-2U}\bar h_{ij}.
\end{equation}
For a conformal rescaling in three dimensions one has
\begin{equation}
\begin{split}
  \mathrm{Ric}(h)_{ij}
  =
  \mathrm{Ric}(\bar h)_{ij}
  + \bar\nabla_i \bar\nabla_j U
  + \partial_iU \partial_jU \\
  + \bigl(\bar\Delta U - |\bar\nabla U|_{\bar h}^2\bigr)\bar h_{ij},
\end{split}
\end{equation}
while for a scalar,
\begin{equation}
  \Delta_h f = e^{2U}\bigl(\bar\Delta f - \bar\nabla U \cdot \bar\nabla f\bigr),
  \qquad
  |Df|_h^2 = e^{2U} |\bar\nabla f|_{\bar h}^2.
\end{equation}
Applying the scalar identity to $f=U$ and using the first static equation gives
\begin{equation}
  \bar\Delta U = - \Lambda e^{-2U}.
\end{equation}

For the Hessian of $U$ one also has
\begin{equation}
  D_iD_jU
  =
  \bar\nabla_i\bar\nabla_jU
  + 2\partial_iU\partial_jU
  - |\bar\nabla U|_{\bar h}^2 \bar h_{ij}.
\end{equation}
Substituting these identities into the second static equation yields
\begin{equation}
  \mathrm{Ric}(\bar h)_{ij}
  =
  2\partial_iU\partial_jU
  + 2\Lambda e^{-2U}\bar h_{ij}.
\end{equation}

\subsection{Axially symmetric three-metric}

Let
\begin{equation}
  \bar h = q_{AB} dx^A dx^B + W^2 d\phi^2,
  \qquad A,B \in \{\rho,z\},
\end{equation}
with all fields independent of $\phi$. Then
\begin{align}
  \mathrm{Ric}(\bar h)_{AB} &= \mathrm{Ric}(q)_{AB} - W^{-1}\nabla_A\nabla_BW, \\
  \mathrm{Ric}(\bar h)_{\phi\phi} &= -W \Delta_q W, \\
  \bar\Delta U &= \Delta_q U + W^{-1}\nabla W \cdot \nabla U.
\end{align}
Combining these identities with the conformally rescaled Einstein equations gives
\begin{align}
  \Delta_q U + W^{-1}\nabla W \cdot \nabla U &= - \Lambda e^{-2U}, \\
  \Delta_q W &= - 2 \Lambda e^{-2U} W, \\
  \mathrm{Ric}(q)_{AB}
    &= W^{-1}\nabla_A\nabla_BW
    + 2\nabla_AU\nabla_BU \nonumber \\
    &\quad
    + 2\Lambda e^{-2U} q_{AB}.
\end{align}

\subsection{Conformal coordinates on the orbit space}

If
\begin{equation}
  q_{AB}dx^A dx^B = e^{2\nu}(d\rho^2 + dz^2),
\end{equation}
then
\begin{align}
  \Delta_q f &= e^{-2\nu}(f_{,\rho\rho} + f_{,zz}), \\
  \mathrm{Ric}(q)_{AB}
    &= -(\nu_{,\rho\rho}+\nu_{,zz})\delta_{AB}.
\end{align}
The mixed and traceless combinations of the tensor equation give
\begin{align}
  \nabla_\rho\nabla_zW &= W_{,\rho z} - \nu_{,z}W_{,\rho} - \nu_{,\rho}W_{,z}, \\
  \nabla_\rho\nabla_\rho W - \nabla_z\nabla_z W
    &= W_{,\rho\rho} - W_{,zz} - 2\nu_{,\rho}W_{,\rho} + 2\nu_{,z}W_{,z},
\end{align}
from which \eqref{eq:nu-first-rho} and \eqref{eq:nu-first-z} follow immediately.

\end{document}